\documentclass[runningheads]{svmult}

\usepackage{makeidx}   
\usepackage{graphicx}  
\usepackage{subeqnar}  
\usepackage{multicol}  
\usepackage{physprbb}  
\makeindex             

\begin{document}
\title*{Rotation-Induced Mixing in Red Giant Stars}
\toctitle{Rotational-Induced Mixing in Red Giant Stars}
\titlerunning{Rotation-Induced Mixing in Red Giant Stars}
\author{Ana Palacios\inst{1}
\and Corinne Charbonnel\inst{2,3}
\and Suzanne Talon\inst{4}
\and Lionel Siess\inst{1}
}
\authorrunning{A. Palacios et al.}
\institute{IAA-Universit\'{e} Libre de Bruxelles, CP-226 Bd du Triomphe,
  B-1050 Brussels, Belgium,
\and Geneva Observatory, 51 chemin des Maillettes, CH-1290 Sauverny, Switzerland
\and LA2T-OMP, UMR 5572, 14 av. E. Belin, F-31400 Toulouse, France
\and D\'epartement de Physique, Universit\'e de Montr\'eal, Montr\'eal PQ H3C 3J7,
 Canada}

\maketitle              

\section{Introduction}
Red giant stars, both in the field and in globular clusters, present
abundance anomalies that can not be explained by standard stellar evolution
models. Some of these peculiarities, such as the decline of $^{12}{\rm
  C}/^{13}{\rm C}$, and that of Li and
$^{12}{\rm C}$ surface abundances for stars more luminous than the
\emph{bump}, clearly point towards the existence of extra-mixing processes
at play inside the stars, the nature of which remains unclear. Rotation
has often been invoked as a possible source for mixing inside Red Giant
Branch (RGB) stars (\cite{SM79},\cite{CC95},\cite{DT00}). In this
framework, we present the first fully consistent computations of
rotating low mass and low metallicity stars from the Zero Age Main Sequence
(ZAMS) to the upper RGB.

\section{Physics of the Models}
We present three models of a 0.85 ${\rm M}_{\odot}$, Z = $10^{-5}$ star.
Model \textbf{A} is a standard non-rotating model.  Model \textbf{B} is a
slowly rotating model with $\upsilon_{\rm ZAMS} = 5~{\rm km}.{\rm s}^{-1}$,
undergoing no magnetic braking, and for which we assumed a solid body
rotating convective envelope (CE) ($\Omega_{\rm CE} = {\rm cst}$) during
the whole evolution.  Model \textbf{C} is an initially rapidly rotating
model with $\upsilon_{\rm ZAMS} = 110~{\rm km}.{\rm s}^{-1}$, which
undergoes magnetic braking ($\upsilon_{\rm TO} = 3~{\rm km}.{\rm
s}^{-1}$), and for which we assumed uniform specific angular momentum in
the CE ($\Omega_{\rm CE} \propto {\rm r}^{-2}$).  In our rotating models,
we compute the transport of both angular momentum and chemicals by
meridional circulation and shear-induced turbulence from the ZAMS on,
according to \cite{MZ98}. We use the new prescription for the horizontal
turbulent viscosity given by \cite{MPZ04}.

\section{Main Results}

Rotational mixing does not affect significantly the stellar structure
(${\rm L}_{bump}$ is the same in models \textbf{A}, \textbf{B} and
\textbf{C}), but leads to larger abundance variations on the lower RGB
associated with a deeper first dredge-up (Fig. 1b).

When solid body rotation is assumed in the CE (model \textbf{B}), the
degree of differential rotation at its base is too low to trigger efficient
shear-induced turbulence between the outer part of the hydrogen burning shell
(HBS) and the CE (solid lines in Fig. 1a). 
On the contrary, in our model \textbf{C} the differential rotation of the
CE ensures the onset of turbulence in the contiguous radiative region,
and the CE is connected to the outer HBS through shear mixing (dotted lines in
Fig. 1a). 
This confirms the conjecture by \cite{SM79}, \cite{SP00} and \cite{DT00}
that the shear-mixing efficiency is enhanced in models with differentially rotating CE.

In our most favorable case (model \textbf{C}), the maximum value of the
diffusion coefficient in the outer part of the HBS (where abundances of Li,
C and N present large variations) is $10^{5}~{\rm cm}^2.{\rm s}^{-1}$
(Fig. 1a), far from the $ 4~10^{8}~{\rm cm}^2.{\rm s}^{-1}$ value that
seems to be necessary to reproduce the observations according to
\cite{DV03}.  As a result none of our rotating models can reproduce the
observed patterns emphasized by \cite{Gratton} (Fig. 1b).

\emph{Rotation remains the best candidate for extra-mixing in
RGB stars. The present modelling of the rotational mixing is however
still incomplete, and agreement between self-consistent models and observations might be
achieved by improving the description of the hydrodynamics related to rotation.} 

\begin{figure}[t]
\includegraphics[width = 6.5cm,height =.35\textheight,angle=-90]{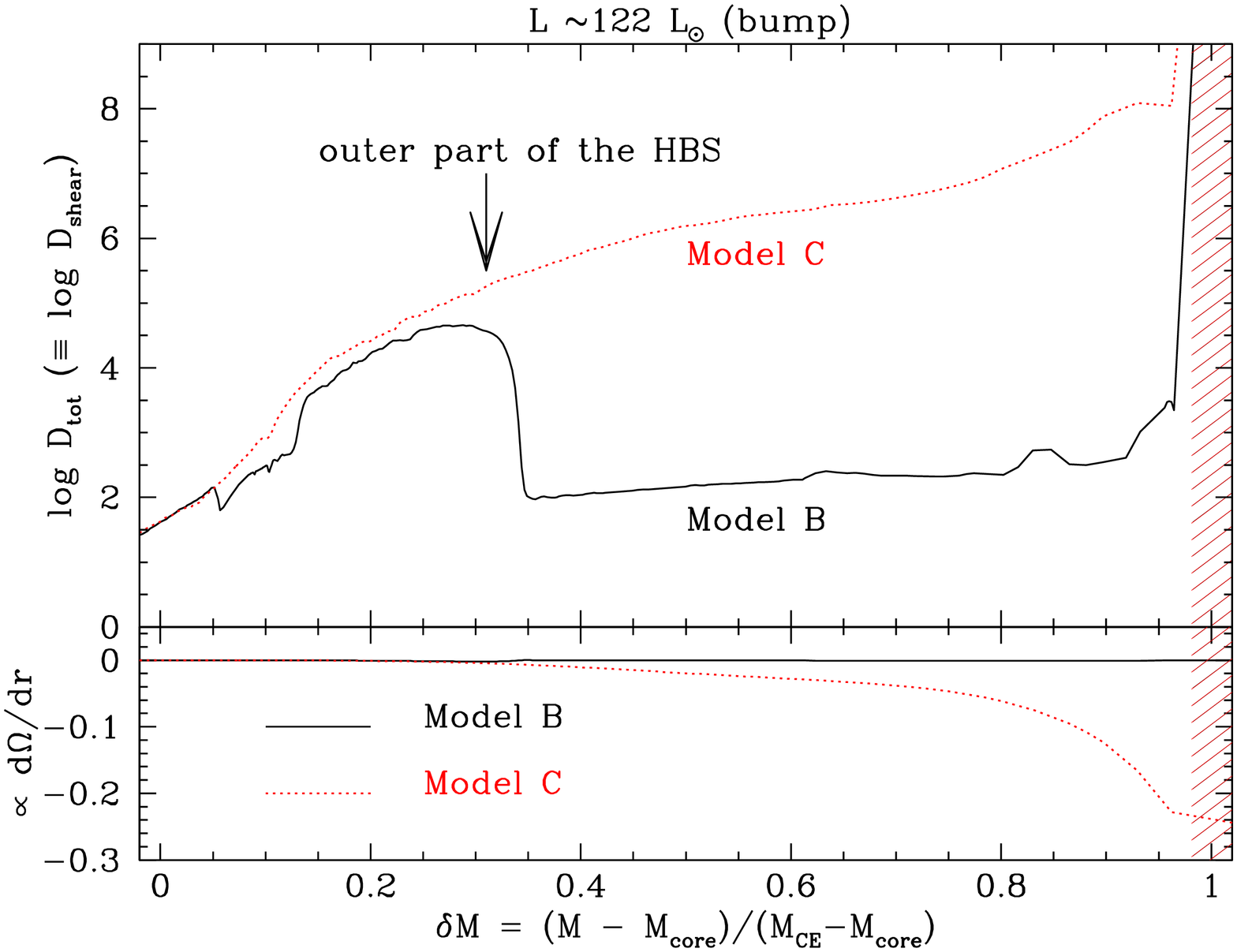}%
\includegraphics[width = 6.5cm,height =.35\textheight,angle=-90]{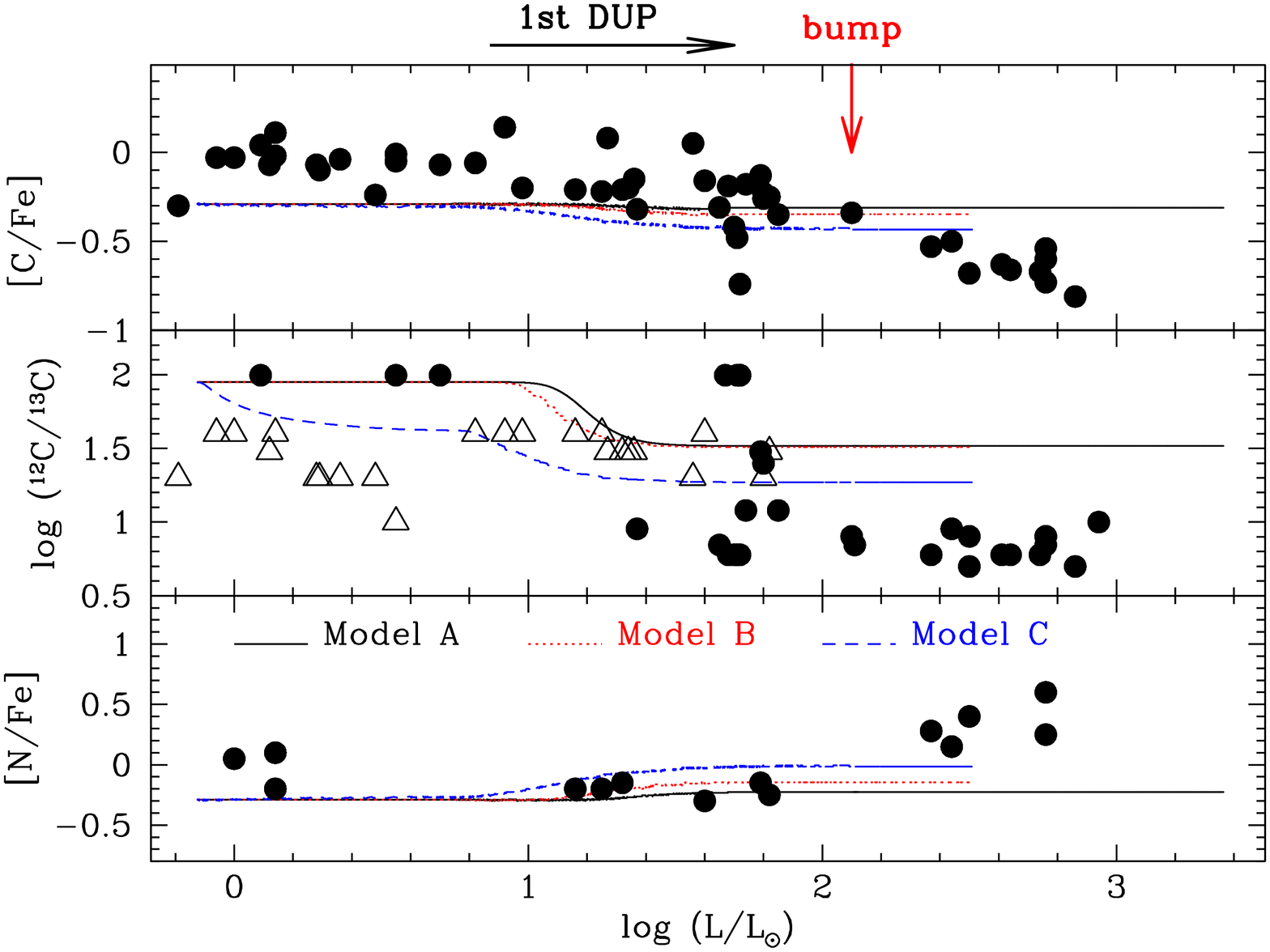}
\caption[]{(\textbf{a}) ({\it left}) Profiles at the bump, of the total diffusion
  coefficient (top) and of the degree of differential rotation (bottom) for model
  \textbf{B} ({\it solid lines}) and model
  \textbf{C} ({\it dotted lines}). Hatched regions correspond to the CE. (\textbf{b}) ({\it right}) Comparison of our models
  with observations ([4]). Triangles are lower limits. Dots are actual values.}
\label{eps1}
\end{figure}

%

\end{document}